# Modulation induced splitting of the magnetic resonance

Rita Behera and Swarupananda Pradhan[*]

Laser and Plasma Technology division, Bhabha Atomic Research Centre, Trombay, Mumbai-400085, India

and Homi Bhabha National Institute, Department of Atomic Energy, Mumbai-400094, India

[*]Correspondence address: spradhan@barc.gov.in and pradhans75@gmail.com

**Abstract**

The splitting of magnetic resonance induced by a linearly polarized frequency comb is presented for Hanle experimental configuration. The experiment is carried out with thermal Rubidium atoms in an anti-relaxation coated glass cell. The observed splitting corresponds to steady-state behaviour of the system over an integrated time scale that is longer than the modulation frequency. The splitting is only observed for dark resonances, with left and right circularly polarized light simultaneously coupled to the atomic system. The missing alternate split components at single photon resonance is consistent with the theoretical calculation. The different components of the split profile have distinct response to the direction of the transverse magnetic field. The response is captured by the theoretical calculation and the underlying physical mechanism is discussed. These studies will be useful in metrology and advancement of vector atomic magnetometer.

**Key words:** magnetic resonance, quantum interference, frequency modulation, transverse magnetic field, frequency comb, single photon resonance

**Introduction:**

The ubiquitous magnetic field carries vital information on the associated physical processes and thus, its measurement is an important part of the contemporary science [1-5]. Apart from fundamental interest, the prospect of highly sensitive magnetometers in a variety of applications has led to exceptional growth in the associated research area [6-8]. The intriguing aspect of atomic interaction with photon gets augmented near zero magnetic field due to interplay of quantum interference, optical pumping and level crossing assisted population redistribution. The semi-classical density-matrix based calculation is a prominent tool for extracting the underlying physical mechanism behind the observations [1, 9-19]. In general, a comprehensive knowledge of the laser atom interaction process is instrumental for the advancement of modern science and atomic devices. The zero-field magnetic resonance, realized through Hanle configuration is central to many of the leading magnetometry technique. The Hanle kind of magnetic resonance in an atomic system is realized by scanning the magnetic field along the laser propagation direction, while keeping the laser frequency close to or away from an atomic resonance.

One variant of Hanle method uses modulated light field for study of magnetic resonance [9-14]. It uses nonlinear magneto-optic rotation for study of atomic spin dynamics near zero magnetic field. The modulation applied to either laser amplitude, frequency or polarization leads to synchronous oscillation of the atomic polarization at the modulated frequency. The oscillating atomic polarization is phase sensitively detected by using or avoiding polarimetric technique. The experimental procedure extracts the oscillation at different harmonics of the modulating frequency by using lock-in amplifier. The amplitude of these oscillations shows resonances as the Larmor's frequency matches with the harmonics of the applied modulation. Resultantly, there is an apparent splitting in the Hanle kind of resonance. However, it may be noted that the process of signal acquisition is different from the conventional Hanle technique. The above configuration is labelled as *FM method* in the subsequent discussion.

We have investigated the splitting of magnetic resonance due to the frequency modulated light field, using a different experimental approach. This method closely resembles with the Hanle configuration and excerpts a different physical attribute (from the *FM method*) of the phenomenon. Similar to the *FM method*, a frequency modulation ($fm$) is applied to the light field. However, an additional low frequency modulation ($mm$) is applied to the magnetic field along the laser propagation direction (longitudinal magnetic field). The transmitted light by the atomic sample is detected in reference with the $mm$ by using a lock-in amplifier. Since the frequency of $mm$ is much slower than $fm$, the oscillations at the frequency of $fm$ gets averaged out in the acquired signal. The signal represents a steady state attribute of the system and is studied as a function of the longitudinal magnetic field. This experimental configuration is termed as *MM method*. In summary, the *FM method* addresses the oscillation in the transmitted light at different harmonics of $fm$, whereas the integrated transmitted light intensity is studied in the *MM method*. These two methods represent two different physical scenarios albeit both originates due to the modulation in the light field and studied as a function of the magnetic field.

The parametric dependence of the signal by *MM method* is studied with respect to the laser polarization, laser detuning, and transverse magnetic field. To the best of our knowledge, these dependences are not addressed in the *FM method* also. The split profiles in *MM method* have shown strong response to these parameters and are discussed in the paper. A simple model based on the density matrix calculation is used to explain the primary feature of the experimental observations.

**Experimental set-up:**

The experiment is carried out with a vertical cavity surface emitting diode laser (VCSEL) tuned to the Rubidium (Rb) D1 line at 795 nm. The laser beam has ~100 μW power with a knife edge width of ~4.5mm, and linewidth <100MHz. The laser frequency is modulated at a frequency $\omega_m$ with an amplitude $A_m$ (12 kHz and ~ 1.63 GHz respectively, unless specified). The primary objective of this modulation is to generate a frequency comb for study of its effect on the Hanle resonance. It is also used for generating an error signal for laser frequency stabilization.

The schematic diagram of the experimental set-up is shown in Fig.-1(A). A part of the laser beam (~10%) is split by the use of a half-wave plate (HWP) and a polarization beam splitter cube (PBS). This beam (in the reflected port of the PBS) is passed through a Rb vapor cell and detected by a photodiode (PD1). An oscillator (Oscillator1) with a frequency $\omega_m$ is used to modulate the injection current of the VCSEL. The PD1 signal is phase sensitively detected by a lock-in amplifier (Lock-in Amp1) in reference with $\omega_m$ to generate an error signal for laser frequency stabilization [8, 14, 20]. A servo loop locks the laser frequency to the zero crossing of the error signal at Rb-85 $F = 3 \rightarrow F' = 2,3$ transition (unless specified) with a stability of ~ <50 MHz. The lock position corresponds to a frequency detuning (Δ) ~+25 × natural line width (Γ) from the Rb85 $F = 3 \rightarrow F' = 2$ transition. The change in the reference lock point leads to frequency stabilization at different offset position of the error signal and is used for study related to Δ dependence of the signal.

The linearly polarized transmitted beam by the PBS, with polarization axis in the x-direction is used for the experiment. A quarter wave plate (QWP) is used after the PBS to change the polarization state of the light field for study related to the laser polarization dependence of the signal. The beam is passed through an anti-relaxation (AR) coated experimental cell at room temperature, containing Rb atoms with natural isotopic composition (Rb2). The AR coated cell preserves the ground state Zeeman coherence against several wall collisions [19, 21]. A solenoidal coil and two set of rectangular coils are used for controlling the magnetic field along and transverse to the laser propagation direction respectively. The utilized electromagnets are spectroscopically calibrated with respect to the coherent population trapping (CPT) signal, generated between the two lower hyperfine levels [20, 22]. All the quoted magnetic field (Bx, By, Bz) in this article are with reference to the zero-magnetic field derived from the splitting of the CPT resonance. The ambient magnetic field is shielded by enclosing the vapor cell along with the electromagnets in several layers of mu-metal sheets.

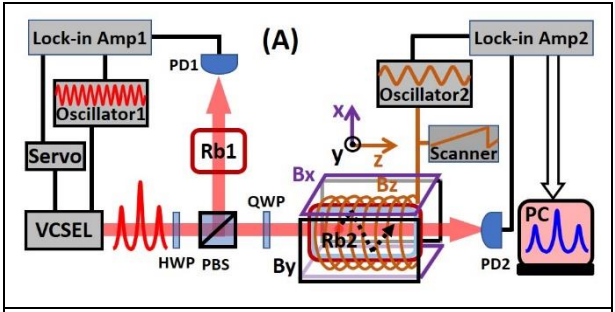

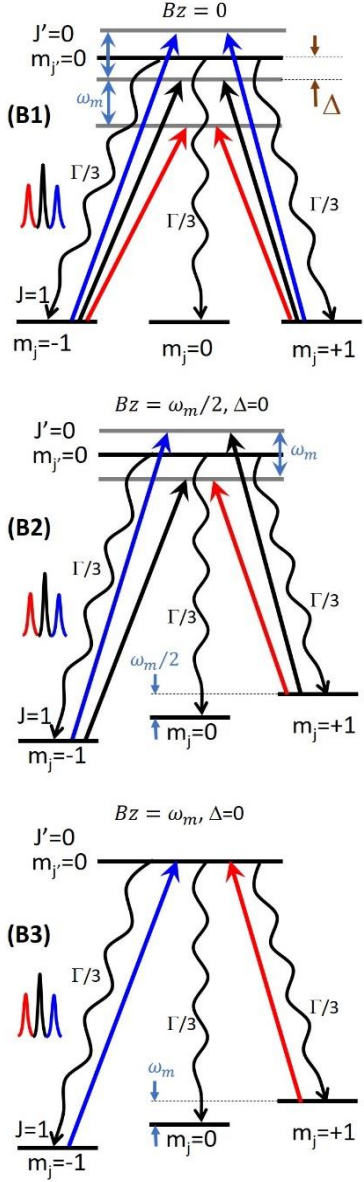

**Fig-1 (A):** Schematic diagram of the experimental set-up. A part of the frequency modulated laser beam after passing through a Rb vapor cell (Rb1), is phase sensitively detected with respect to the modulation applied to the laser field and is used for laser frequency stabilization. The other part of the beam after interaction with Rb atoms in an AR coated vapor cell (Rb2), is phase sensitively detected with respect to modulation applied to the Bz field for study of magnetic resonances. The laser polarization axis is along x-direction. (Please see text for details)

**(B):** Model used for the calculation. The $J = 1 \rightarrow J' = 0$ transition is coupled with a frequency modulated linearly polarized light field. The coupling of the carrier (black) and ±1 side mode is illustrated for different experimental conditions. The higher side modes are not shown here, though they are used in the calculation. This simplified model is sufficient to explain the central part of the experimental observation and helpful for visualizing the underlying physical mechanism.

The scanning of the Bz field is done by a ramp generator (Scanner) and an oscillating field is imposed by an oscillator (Oscillator2) as shown in Fg.1(A). The transmitted light through Rb2 is phase sensitively detected using a lock-in amplifier (Lock-in Amp2) in reference with the Oscillator2 and is termed as MMz signal. The oscillating magnetic field has a frequency of 55Hz with an amplitude ~200 nT. The frequency of modulation applied to the magnetic field is kept slower than $\omega_m$ to isolate the MMz signal from the oscillation at $\omega_m$ and its harmonics present in the amplitude of the light field. This is in contrast to the *FM method* [9-14], where the investigated signal exclusively represents the oscillation at $\omega_m$ and its harmonics. Thus, the MMz signal presented in this work is technically different compared to all the prior investigations. The *MM method* has a technical advantage compared to the conventional *FM method* as the contribution of the residual amplitude modulation to the signal profile is inherently circumvented.

### Theoretical outline:

The objective of the theoretical calculation is to capture the observed splitting of the magnetic resonance and their response to the transverse magnetic field. The calculation is carried out by using some of the basic blocks available in the "Atomic Density Matrix" package [1, 17, 18, 23]. The density matrix formalism for calculating the response of an atomic ensemble near zero magnetic field, while interacting with a light field has been described by several group [1, 9-13, 15-18]. For completeness, the model, associated interaction, relaxation and procedure for calculation (in consistency with the *MM Method*) are briefly discussed. The level diagram of an atomic transition ($J = 1 \to J' = 0$) coupled with a linearly polarized light field for different experimental condition is shown in Fig-1(B). This is an ideal system for study of ground state Zeeman coherence and associated physical processes. The mf=0 ground state is not coupled to any light field and mimics the uncoupled ground hyperfine level in the realistic alkali atoms. The total Hamiltonian corresponding to the interaction of a light field ($\mathcal{E}$) with the atom in presence of a magnetic field ($\boldsymbol{B}$) is given by $H = H_0 - \boldsymbol{d}.\boldsymbol{\mathcal{E}} + \boldsymbol{\mu}.\boldsymbol{B}$.

The input light field propagating along z axis ($E_0(\cos\epsilon\ \hat{x} + i\ \sin\epsilon\ \hat{y})\cos\omega t$) is frequency modulated by a sinusoidal wave $A_m \cos(\omega_m t)$. The resultant frequency modulated field (taking care of accumulated phase) can be represented as $E_0(\cos\epsilon\ \hat{x} + i\ \sin\epsilon\ \hat{y})\cos\left(\omega t + \frac{A_m \cos(\omega_m t)}{\omega_m}\right)$. The decay matrix for the associated spontaneous decay rate ($\Gamma$) and collisional decay rate ($\gamma_{ex}$) are calculated from Lindbald super operator $\mathcal{L}_d \rho = \left[C_{mn}\rho C_{nm} - \frac{1}{2}(\rho C_{nm}C_{mn} + C_{nm}C_{mn}\rho)\right] \times (\Gamma/3, \gamma_{ex})$, where $C_{mn} = |m\rangle\langle n|$ is the quantum jump operator. The collisional relaxation among all the ground Zeeman states is considered with equal probability. The relaxation and repopulation associated with ground state coherence relaxation rate ($\gamma_c$) are is given by $\mathcal{L}_t \rho = -\frac{1}{2}\{R, \rho\} + \Lambda$. The evolution of the density matrix for this system is obtained by solving the Lindbald master equation $\frac{\partial \rho}{\partial t} = -\frac{i}{\hbar}[H, \rho] + \mathcal{L}_d \rho + \mathcal{L}_t \rho$. The entire notations utilized in the above description have their customary meaning.

The influence of transverse magnetic field to the atomic dynamics can be incorporated by transforming the electric field vector (of the light field) along the direction of magnetic field [24, 25]. However, coupling between the Zeeman ground state by the transverse field can be equally used. This approach preserves the quantization axis along the laser propagation direction [18, 23]. We have utilized the later approach for calculation of the signal profile. The selection rule $\Delta m_j = \pm 1$, for magnetic coupling between ground level Zeeman states is used. The coupling strength is decided by the amplitude of corresponding magnetic field.

The amplitude of the temporal oscillation of density matrix element, at different harmonics of $\omega_m$ is calculated in the *FM method*. In the *MM method*, we integrate this oscillatory function over several time periods (60 periods are used for calculation) of the applied modulation. As the integrated signal is steady with respect to time, it represents a kind of steady state behaviour of the system. The derivative of the above integrated signal is compared with the experimental data owing to the use of phase sensitive technique in the experiment. The integrated signal is calculated as a function of the Bz field. The calculation is carried out for linearly polarized light with polarization axis along x-direction. The values of the parameters used for calculation are kept close to the experimental values as given below.

The calculations are carried out with $\Gamma$ =5.6 MHz, laser detuning ($\Delta$) = +25× $\Gamma$ unless specified, Rabi frequency = 1 MHz, $\gamma_{ex}$ = 500Hz, $\gamma_c$ = 500Hz, $\omega_m$ = 12 kHz, $A_m$ = 100 MHz, and integration time = 2 ms. The increase in the value of $\gamma_{ex}, \gamma_c$, or Rabi frequency is accompanied by the broadening of the line shape at different rates. The values of $\gamma_{ex}, \gamma_c$ and $A_m$ are selected to bring the calculated and experimental line shape closer to each other. The other parameters are same as the experimental values. The lower value of $\gamma_c$ is critical for realizing narrow resonance width and is achieved by the use of AR coated atomic vapor cell. The $\gamma_c$ can be equally improved by the use of Buffer gas filled cell. However, it is accompanied with very large homogeneous (collisional) broadening that is detrimental for the phenomena of interest. The reason for using smaller value of $A_m$ is discussed in the following section.

**Results and discussions:**

The prominent physical mechanisms behind the Hanle kind of resonance are optical pumping followed by Zeeman redistribution and quantum interference between excitation pathways. The various experimental parameters like laser polarization, laser intensity, magnetic field direction & amplitude, atomic properties, and inhabitant composition in the cell dictate the relative domination of these effects [8-27]. The signal acquired through the *MM method* with linearly polarized light shows splitting of the Hanle kind of resonance ($\Delta$=+25× $\Gamma$ from Rb85 $F = 3 \to F' = 2$ transition), as shown in Fig.-2. The dispersive signal profile is due to the use of phase sensitive detection technique. The positive and negative slope of the signal corresponds to enhanced transmission and absorption respectively. Any unintentional alternation of the signal profile due to the Bz field modulation is verified by changing the frequency and amplitude of the magnetic modulation form 13 Hz to 79 Hz and 10nT to 400 nT respectively. The features of the MMz signal profile remained intact except for the changes in the amplitude and associated noise level. Thus, the magnetic field modulation neither introduces any extra broadening nor alters the MMz signal profile. The Bz field separation between the neighbouring split components is found to be ~1108 nT ($\omega_{L01}$) for $\omega_m$=12 kHz, as shown in Fig.2. The gyromagnetic ratio for Rb-85 ground level is ~4.7 Hz/nT that gives the theoretical value of $\omega_m/2$ to be ~1276 nT [28]. The close value of the $\omega_{L01}$ with the theoretical value for $\omega_m/2$ indicates that the resonance occurs as the Larmor's frequency approaches a harmonic of $\omega_m/2$. Similar splitting is also observed for laser locked at a detuned position from the Rb87 $F = 2 \rightarrow F' = 1$ transition. The gyromagnetic ratio for Rb-87 ground level is 7 Hz/nT that gives the theoretical value of $\omega_m/2$ (for $\omega_m$=12 kHz) to be ~857 nT. The measured separation between nearby split component is ~748 nT ($\omega_{L02}$) that is also close to the corresponding theoretical value. The difference of $\omega_{L01}$ and $\omega_{L02}$ from the theoretical values reflects the accuracy of the magnetic field calibration.

The observed splitting in the *MM method* is due to enhanced macroscopic atomic polarization (steady state) as Larmor's frequency became resonant with any harmonics of $\omega_m/2$. There is a resemblance with the *FM method*, where the oscillations (at the harmonics of $\omega_m/2$) of the atomic polarization are resonantly enhanced under identical condition. The splitting is predominantly observed for linearly polarized light and its polarization dependence provides information on the underlying mechanism. A single enhanced absorption profile (without splitting) is observed for either left or right circularly polarized light as shown in Fig.-2. In the current experimental conditions, it is impossible to realize a common level excitation for a pure circularly polarized light field. Thus, it is the optical pumping followed by Zeeman redistribution rather than quantum interference that contributes to the signal for pure circularly polarized light field. This mechanism is consistent with the enhanced absorption observed for the circularly polarized light. It reveals that the signal originating due to optical pumping followed by Zeeman redistribution does not doesn't show any split profile in *MM method*. This is in contrast to *FM method*, where splitting is observed for circularly polarized light.

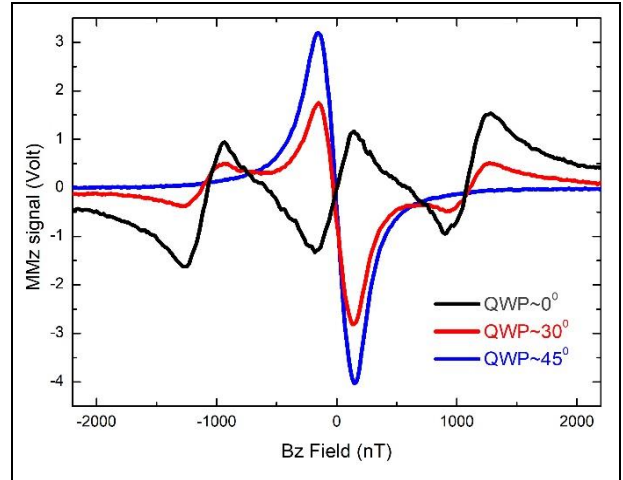

**Fig-2:** Polarization dependence of the MMz signal. The split components show enhanced transmission for linearly polarized light (black curve). For elliptically polarized light (QWP ~30⁰), the central split component gets transformed to enhanced absorption (red curve). A single enhance absorption profile is observed for pure circularly polarized light (blue curve). The experiments are carried out in absence of transverse magnetic fields.

For the linearly polarized light field, quantum interference plays role in the atomic dynamics near zero magnetic field as shown in Fig.-1(B). The dark and bright resonances (enhanced transmission and absorption) are generally associated with $F_g \rightarrow F_e \leq F_g$ and $F_g \rightarrow F_e > F_g$ class of transition respectively [16, 18, 20, 24, 25]. In the *MM method* with linearly polarized light, enhanced transmission is observed for all Rb-85 and Rb-87 D1 transition with $F_e \leq F_g$ except for Rb-85 $F = 2 \rightarrow F' = 2$. The contradiction is due its spectral overlap with the Rb-85 $F = 2 \rightarrow F' = 3$ transition that exhibits bright resonance. These dark resonances are found to be transformed to bright resonance as the laser polarization is changed from linear to circular (Fig.-2). None of the bright resonances ($F_g \rightarrow F_e > F_g$) of Rb-85 and 87 D1 transition showed splitting of the magnetic resonance irrespective of the laser polarization. Similarly, splitting is not observed for pure circularly polarized light as all the transitions shows enhanced absorption. Thus, the observation of splitting in the *MM method* is facilitated by simultaneous coupling of both left (σ-) and right (σ+) circularly polarized light to a dark resonance. We couldn't observe similar splitting of the Hanle resonance in buffer gas (@25 Torr Nitrogen gas) environment. This is due to very large homogeneous (collisional) broadening (~few GHz) that leads to spectral overlap among the transitions. It is consistent with the above discussed anomaly for the Rb-85 $F = 2 \rightarrow F' = 2$ transition where splitting is not observed. The increase in the homogeneous width also broadens the split components as discussed in the theoretical section and obscures the observation of the splitting. Thus, AR coated cell is not only required for improving the Zeeman coherence time but also an ideal system for observation of the split profile in the *MM Method*.

The modulation of the VCSEL injection current by $A_m \cos(\omega_m t)$ leads to the generation of a frequency comb with side bands spaced at $\omega_m$. The intensity of the nth side-band ($I_n$) is $\propto |J_n(A_m)|^2$, where $J_n(A_m)$ is a $n^{th}$-order ordinary Bessel function. The σ- and σ+ components of these side bands constitute CPT states as the Larmor's frequency becomes resonant with the harmonics of $\omega_m/2$. The factor ½ arises due to the frequency shifting between $m_j = -1$ and $m_j = +1$ (participating in CPT, Fig-1B) at twice rate of the associated Larmor's frequency. Thus, the separation and relative amplitude of the split components of the magnetic resonance corresponds to $\omega_m/2$ and $\propto |J_n(A_m)|^2$ respectively, and is consistently observed in Fig-3 (A). As $A_m$ is increased, CPT states at higher Larmor's frequency are observed due to coupling of more power to the higher side modes. The derivative of the calculated signal profile is shown in Fig.-3B that can be compared with the experimental profile. The calculated split profile consistently explains the experimental observation under different conditions and is discussed in the following part of this article. The success of this simplified model (shown in Fig.-1B) proves the ground state Zeeman coherence as the mechanism behind the observation.

The split components are categories in to three group, the central component near Bz=0, components $n \times \omega_m/2$ (where $n$ is an odd integer), and components at multiple of $\omega_m$. The central split component at Bz=0 is observed for both resonant (Δ=0) and off resonant (Δ≠0) light field as shown in Fig.-3(B). The corresponding coupling diagram (for carrier and ±1 side modes) is shown in Fig-1(B1), where the σ+ and σ- component of each mode constitutes a lambda system. The higher order modes also form similar lambda system at different detuned position and are not shown in Fig.-1(B1). These lambda systems have different contribution to the signal depending on the value of Δ and $A_m$. In contrast, the σ+ and σ- component from two different laser modes constitutes the CPT state for other components of the split profile. The mf=0 ground state in Fig.-1B resembles to the non-coupled ground hyperfine level in the realistic atomic system. The population of the mf=0 state is reduced due to the creation of CPT state (leading to diminishing optical pumping) at each of the split components.

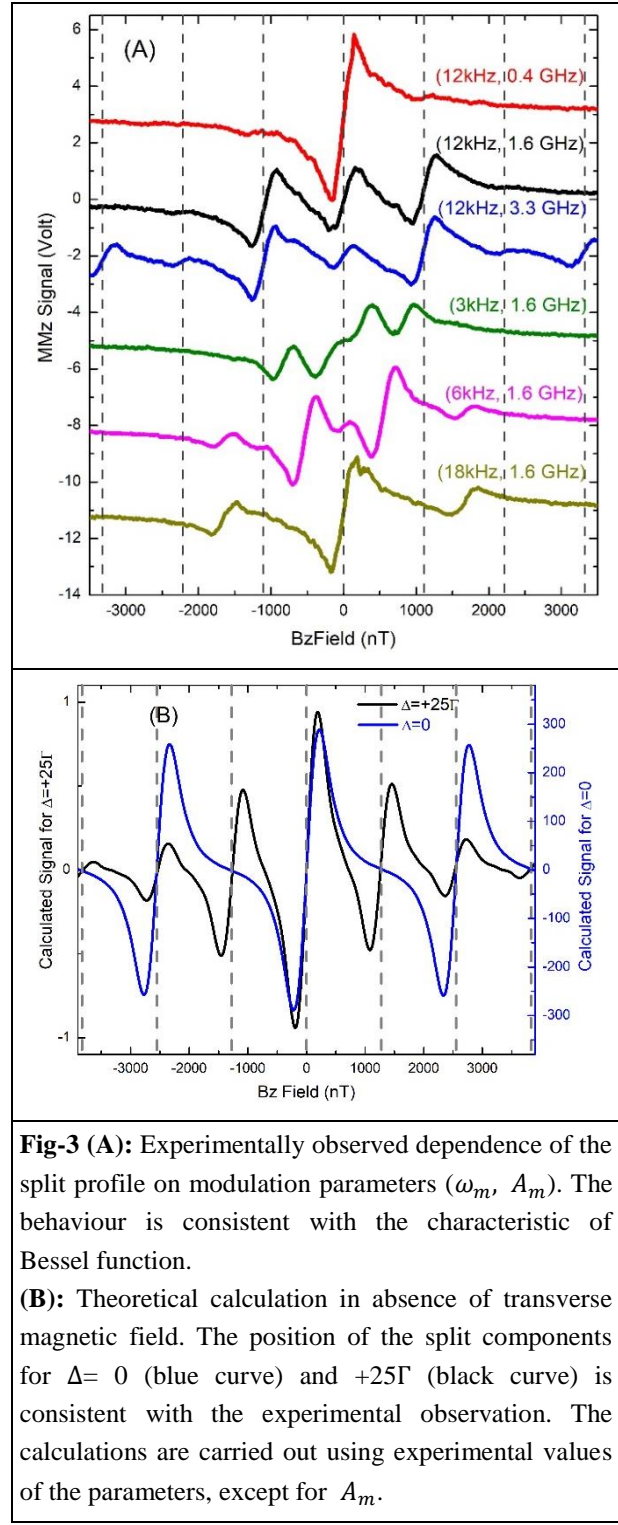

**Fig-3 (A):** Experimentally observed dependence of the split profile on modulation parameters ($\omega_m$, $A_m$). The behaviour is consistent with the characteristic of Bessel function.

**(B):** Theoretical calculation in absence of transverse magnetic field. The position of the split components for Δ= 0 (blue curve) and +25Γ (black curve) is consistent with the experimental observation. The calculations are carried out using experimental values of the parameters, except for $A_m$.

The calculated split components at $n \times \omega_m/2$ are not observed for laser frequency resonant ($\Delta=0$) with the atomic transition. The phase of $\pm$nth side modes around the carrier frequency follows the relationship $J_{-n}(A_m) = (-1)^n J_n(A_m)$. Thus, the +1 and -1 side modes have opposite phase with each other. For this specific case (Bz=$\omega_m$/2 and $\Delta=0$), two symmetrically placed (with respect to the excited level) lambda systems forms a close loop excitation as shown in Fig.1(B2). Since one of the field is in opposite phase with respect to the other fields (due to the above relationship), the quantum interference is destroyed in this symmetric close loop excitation. Similar physical mechanism leads to the disappearance of the split components at all $n \times \omega_m/2$ for $\Delta=0$. For $\Delta \neq 0$, one of the lambda systems becomes dominant over the other in the close loop excitation as the above symmetry is broken. Consequently, the split components at $n \times \omega_m/2$ are observed for $\Delta \neq 0$. The coupling scheme for Bz=$\omega_m$ and $\Delta=0$ is shown in Fig.-1(B3), where the $\sigma$+ component of the +1 side mode forms a lambda system with the $\sigma$- component of the -1 side mode. It is evident from the coupling diagram that the split component at multiple of $\omega_m$ is observed for both on and off resonant light field.

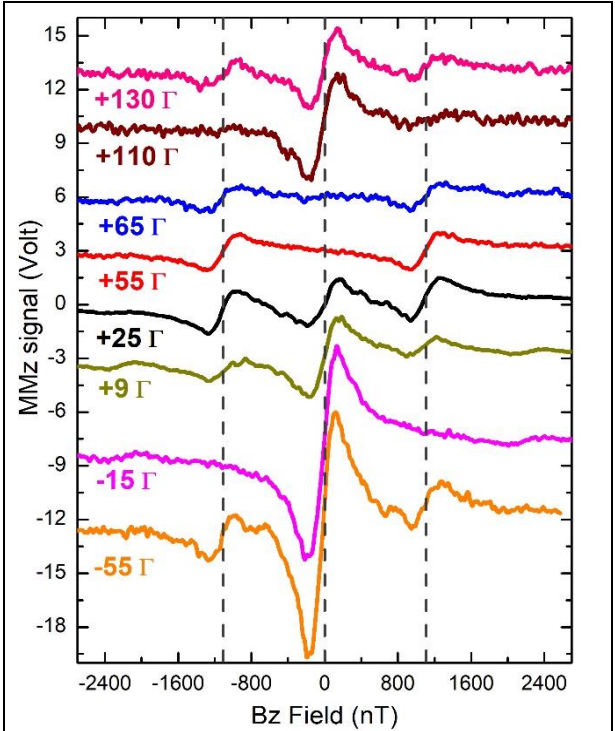

**Fig-4:** The amplitude of the different split components of the MMz signal profile shows dependency on the laser detuning. The detuning values ($\Delta$) are with respect to Rb85 $F = 3 \rightarrow F' = 2$ transition.

The experimentally observed detuning dependence of the split components is shown in Fig.-4. The primary feature of the MMz profiles are consistent with the calculated signal profile for $\Delta$=+25$\Gamma$. The experimental signal profile at $\Delta$=-15$\Gamma$ and +110$\Gamma$ exhibit similar behaviour to the calculated profile for $\Delta$=0, where the components at $\omega_m$/2 are supressed. However, the resonance position for Rb85 $F = 3 \rightarrow F' = 2$ and 3 transitions are at $\Delta$=0 and +65 $\Gamma$ respectively. It may be noted that the signal profile will have contribution from both of the transitions due to thermal velocity of the atoms. A detail calculation with the realistic atomic system (with thermal averaging) is required to address the discrepancy.

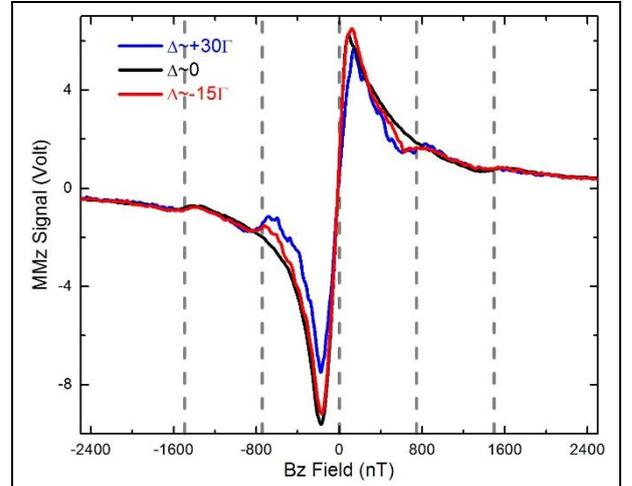

**Fig-5:** Dependence of the split profile for different detuning ($\Delta$) from the Rb87 $F = 2 \rightarrow F' = 1$ transition. The split components at ~ ± 748 nT are not seen for $\Delta$~0 (black curve), whereas the components at ~ ± 1496 nT are present in all the curve, consistent with the calculated profile shown in Fig-3 (B).

The detuning dependence of the MMz signal profile is further studied for Rb-87 $F = 2 \rightarrow F' = 1$ transition coupled with a linearly polarized light. It is an ideal system to study the detuning dependence as there is no overlapping transition. The experiment is carried out with $\omega_m$=12 kHz and $A_m$ ~ 1.63 GHz, same as in Fig.4. The split component at $\omega_m$/2 are not observed for laser tuned to the Rb-87 $F = 2 \rightarrow F' = 1$ transition as shown in Fig.-5. These components appear for off resonant light field and the observation is consistent with the calculated profile shown in Fig.-3(B). The disappearance of the split components at $n \times \omega_m/2$ for $\Delta=0$ can be utilized as a new sub-doppler spectroscopic technique for precise measurement of atomic energy levels while using a broad frequency comb.

The relative amplitude of the split components in Fig.-5 is different from the corresponding signal for Rb-85 atoms shown in Fig.-2, despite using same value of $\omega_m$ and $A_m$. The split components for Rb-87 atoms become more prominent at higher value of $A_m$ and signal profile appears similar to Fig.-2. Thus, the relative amplitude of the split components depends on the details of the transition apart from $A_m$. These details are not incorporated in the model resulting in requirement of a smaller value of $A_m$ to match with the experimental results. However, the utilized simple model demonstrates the generality of the phenomena that will have application in other system like ultracold atoms. It also provides the vital physical processes without going through the involved calculation.

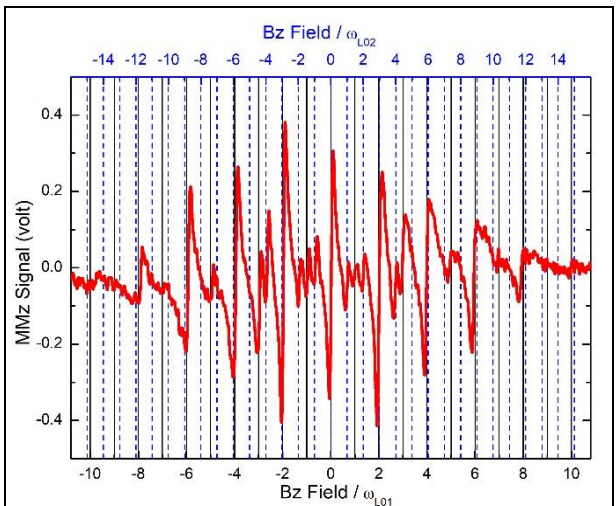

**Fig-6:** The splitting of the magnetic resonance for $\omega_m$=12kHz and $A_m$~6.5 GHz. The black solid line and blue dashed line are marked at an interval of $\omega_{L01}$ (Rb-85) and $\omega_{L02}$ (Rb-87) respectively.

The MMz signal profile shows additional features in AR coated cell for higher amplitude of $A_m$(~6.5 GHz). The higher $A_m$ also broadens the single photon absorption spectrum, similar to the buffer gas filled cell. It is due to generation of a wider frequency comb whereas a large homogeneous broadening was responsible for the buffer filled gas filled cell as has been discussed earlier. The laser frequency is locked to the zero crossing of this broad absorption signal, where the central mode of the frequency comb is expected to be at $\Delta$~+25 $\Gamma$ from the Rb85 $F = 3 \rightarrow F' = 2$ transition. The split components at higher Bz field are prominently observed as shown in Fig.6, despite having a broad single photon resonance. This is consistent with the calculation, where a larger $A_m$ leads to observation of the higher order split components whereas larger homogeneous broadening has detrimental role for observation of the split profile. It may be noted that split profile was not observed for buffer gas filled cell.

The higher side-mode of the light field are close to the Rb87 $F = 2 \rightarrow F' = 2, 1$ transitions at large value of $A_m$. The spacing between split components ($\omega_m/2$) for Rb 85 and Rb 87 atoms are $\omega_{L01}$ and $\omega_{L02}$ respectively. Thus, the split components of the Hanle resonance for Rb-85 and Rb87 atoms appear at different Bz magnetic field in the MMz signal profile (Fig.-6). The split components from both the species are merged together at several Bz field leading to apparent amplified signal, whereas some components have diminished amplitude due to partial overlap with each other. It will be interesting to further explore the experiment to find out the single photon resonance (of the central mode of the frequency comb) for different transition by observing the depleted split components at $\omega_m/2$. This cannot be achieved by conventional sub-doppler spectroscopic technique (like saturation absorption spectroscopy) due to wider spectrum of the frequency comb. A detail investigation in this regard is beyond the scope of this article.

The direction of the ambient magnetic field with respect to the polarization axis of the light field plays a critical role in the dynamics of the laser-atom interaction. The Bx field is parallel to the light polarization axis, whereas the By and Bz field are perpendicular to it. The distinct response of the atomic dynamics to the Bx and By field are reflected in the MMz signal profile as shown in Fig.-7. The quantum interference and optical pumping are the driving mechanism behind the MMz signal profile. The interplay and their dominance depend on the relative amplitude between the magnetic field along and transverse to the polarization axis of the light field. The various conclusions are drawn in the following part of the article, based on the calculated change in the population of mf=0 ground state. The depletion and enhancement in the population of the mf=0 ground state are presumed as the establishment of dark and bright state (due to quantum interference) respectively.

The transverse By field being orthogonal to the laser polarization axis, gets added with the scanning Bz field. The resultant scanning field becomes $\sqrt{B_z^2 + B_y^2}$ and is perpendicular to the laser polarization axis. The position of the split components gets shifted with change in the By field as shown in Fig.-7 (A1) & (B1). The position of the split component will remain unaltered if plotted against $\sqrt{B_z^2 + B_y^2}$ instead of Bz field. This may be realized from the larger shift in the position of the ±1 split components (at $\pm \omega_m/2$) towards the centre as compared to ±2 split components (at $\pm \omega_m$) for the same value of By field. Interestingly, the central component gets attenuated with increase in the amplitude of By field and eventually vanishes (skipped during the scanning) for By~900nT due to the same reason. On further increase in the By field, the ±1 split components merge together and appears as the central resonance (not shown here). In contrary to the central component in absence of the orthogonal field, this apparent central structure at higher By field is originated due to quantum interference involving the σ+ and σ- components from two different modes of the frequency comb. The dark state is established at each of the split component (for different By field value) as has been seen by the depletion (calculated) in the population of the mf=0 ground state.

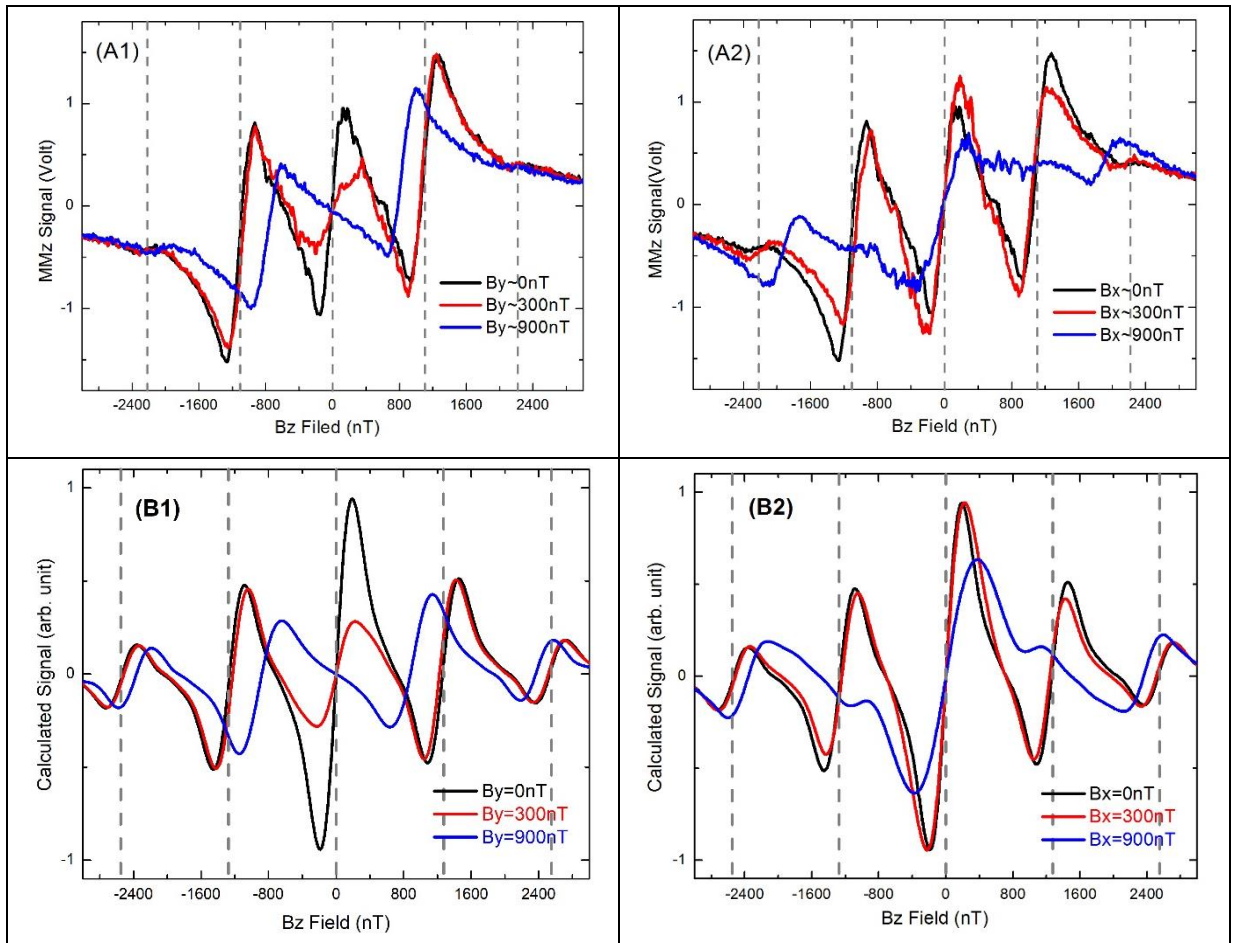

**Fig-7:** Experimental MMz signal (**A1, A2**) and calculated signal (**B1, B2**) profiles for different amplitude of Bx and By field as a function of Bz field. The calculated signal profile remarkably captures the details of experimental observation. The signal profiles are identical for change in the polarity of Bx and By field (not shown here). The experiment is carried out at $\omega_m$=12 kHz and $A_m$ ~ 1.63 GHz, whereas the calculated profiles corresponds to $\omega_m$=12 kHz and $A_m$ ~ 100 MHz.

The transverse Bx magnetic field being along the laser polarization axis, influences the MMz signal profile in a different manner. The relative amplitude of the Bx field as compared to the scanning Bz field, dictates the onset of different physical mechanism. For the Bz scan in the regime Bz >Bx, the Bx field gets added to the scanning Bz field ($\sqrt{B_z^2 + B_x^2}$), in a similar way to the additional By field. Thus, the split components shift towards the centre (Fig.-7 A2 and B2) as the amplitude of Bx field is increased. The split components in this regime behave similar to the additional By field. For the Bz scan regime with Bz < Bx (near the central split component), the dark state is compromised. It is consistently observed by enhanced optical pumping of the atomic population to the mf=0 ground state in the calculated data. Similar mechanism has been reported in different experimental configuration where the dark resonance is transformed to a bright resonance in presence of transverse magnetic field [16, 18, 24, 25]. Since there is a depletion in the population of the mf= ±1 ground state, enhanced transmission of the optical field is observed despite the dark state is disturbed by the Bx field. This mechanism dominates at the central part of the MMz profile till Bz~Bx, leading to broadening of the central resonance with increase the value of Bx field. The calculated population of the mf=0 ground state shows steady increase with amplitude of the Bx field in this regime. The amplitude of the ±1 split components also get reduced apart from shift towards centre, as the amplitude of the Bx field approaches $\omega_m/2$. It is shown in the plot for Bx~±900 nT and proves the detrimental role of the Bx field for the dark state. On further increase in the Bx field, a single board profile is observed without any split components (not shown here). As has been stated, this broad profile is due to optical pumping resulting from the collapse of the dark state. This further establishes the requirement of dark state for observation of the split components in the *MM method*. In summary, the central component gets skipped during scanning of the Bz field as the amplitude of By field is increased, whereas enhanced optical pumping (to the uncoupled mf=0 state) facilitated by the Bx field leads to broadening of the central component. Nevertheless, the side components move towards the centre of the MMz profile as either of Bx or By field is increased. The distinct response of the MMz signal profile to the direction of the transverse magnetic field will be useful for advancement of three axis atomic magnetometer.

## Conclusions:

The splitting of the Hanle resonance by a frequency modulated light field is investigated using *MM method* and its distinctions with the *FM method* are discussed. The integrated atomic polarization gets resonantly enhanced as the Larmor frequency becomes resonant with the harmonics of the applied modulation. This is an additional feature to the oscillating atomic polarization studied in the conventional *FM method*. The splitting of the magnetic resonance for different experimental parameters is studied using a linearly polarized light field. The essential conditions required for observation of splitting in the *MM method* are pointed out. The sharp dependence of the split components (at the odd integral multiple of $\omega_m/2$) on the single photon detuning can have application in metrology infrequency comb. The contradistinctive nature of the split components to the direction of magnetic field, along with the underlying physical processes is presented.

## Acknowledgements:

The authors are thankful to Dr. Shailesh Kumar for supporting the research work.

## Disclosures:

The authors declare no conflicts of interest.